\documentclass[twocolumn,epjc3]{svjour3}
\usepackage{amsmath, amssymb, amsfonts}
\RequirePackage{graphicx}

\journalname{Eur. Phys. J. C}

\begin{document}

\title{First search for dark matter annihilations in the Earth with the IceCube Detector}
\onecolumn
\author{IceCube Collaboration: M.~G.~Aartsen\thanksref{Adelaide}
\and K.~Abraham\thanksref{Munich}
\and M.~Ackermann\thanksref{Zeuthen}
\and J.~Adams\thanksref{Christchurch}
\and J.~A.~Aguilar\thanksref{BrusselsLibre}
\and M.~Ahlers\thanksref{MadisonPAC}
\and M.~Ahrens\thanksref{StockholmOKC}
\and D.~Altmann\thanksref{Erlangen}
\and K.~Andeen\thanksref{Marquette}
\and T.~Anderson\thanksref{PennPhys}
% ...
}
\authorrunning{IceCube Collaboration}
\thankstext{a}{Corresponding authors: jan.lunemann@vub.ac.be, jan.kunnen@vub.ac.be}
\thankstext{b}{Earthquake Research Institute, University of Tokyo, Bunkyo, Tokyo 113-0032, Japan}
\institute{
% ...
Department of Physics, University of Adelaide, Adelaide, 5005, Australia \label{Adelaide}
% ...
\and Universit\'e Libre de Bruxelles, Science Faculty CP230, B-1050 Brussels, Belgium \label{BrusselsLibre}
% ...
\and Dept.~of Physics and Astronomy, University of Canterbury, Private Bag 4800, Christchurch, New Zealand \label{Christchurch}
% ...
\and Erlangen Centre for Astroparticle Physics, Friedrich-Alexander-Universit\"at Erlangen-N\"urnberg, D-91058 Erlangen, Germany \label{Erlangen}
% ...
\and Dept.~of Physics and Wisconsin IceCube Particle Astrophysics Center, University of Wisconsin, Madison, WI 53706, USA \label{MadisonPAC}
% ...
\and Department of Physics, Marquette University, Milwaukee, WI, 53201, USA \label{Marquette}
% ...
\and Physik-department, Technische Universit\"at M\"unchen, D-85748 Garching, Germany \label{Munich}
% ...
\and Oskar Klein Centre and Dept.~of Physics, Stockholm University, SE-10691 Stockholm, Sweden \label{StockholmOKC}
% ...
\and Dept.~of Physics, Pennsylvania State University, University Park, PA 16802, USA \label{PennPhys}
% ...
\and DESY, D-15735 Zeuthen, Germany \label{Zeuthen}
} 
\date{Received: date / Accepted: date}
\maketitle
\twocolumn
\begin{abstract}
We present the results of the first IceCube search for dark matter annihilation in the center of the Earth. Weakly Interacting Massive Particles (WIMPs), candidates for dark matter, can scatter off nuclei inside the Earth and fall below its escape velocity. Over time the captured WIMPs will be accumulated and may eventually self-annihilate. Among the annihilation products only neutrinos can escape from the center of the Earth. Large-scale neutrino telescopes, such as the cubic kilometer IceCube Neutrino Observatory located at the South Pole, can be used to search for such neutrino fluxes.

Data from 327 days of detector livetime during 2011/ 2012 were analyzed. No excess beyond the expected background from atmospheric neutrinos  was detected. The derived upper limits on the annihilation rate of WIMPs in the Earth ($\Gamma_A = 1.12 \cdot 10^{14} ~ \rm{s}^{-1}$ for WIMP masses of 50 GeV annihilating into tau leptons) and the resulting muon flux are an order of magnitude stronger than the limits of the last analysis performed with data from IceCube's predecessor AMANDA. The limits can be translated in terms of a spin-independent WIMP-nucleon cross section. For a WIMP mass of 50 GeV this analysis results in the most restrictive limits achieved with IceCube data.   
\end{abstract}

\section{Introduction}
\label{sec:intro}
A large number of observations, like rotation curves of galaxies and the cosmic microwave background temperature anisotropies, suggests the existence of an unknown component of matter~\cite{bib:Bertone}, commonly referred to as dark matter. However, despite extensive experimental efforts, no constituents of dark matter have been discovered yet. A frequently considered dark matter candidate is a Weakly Interacting Massive Particle~\cite{bib:Steigman}. Different strategies are pursued to search for these particles: at colliders, dark matter particles could be produced~\cite{bib:colliders}, in direct detection experiments, nuclear recoils from  a massive target could be observed~\cite{bib:Xenon100,bib:LUX,bib:SuperCDMS,bib:Cresst}, and indirect detection experiments search for a signal of secondary particles produced by self-annihilating dark matter~\cite{bib:Fermi_DM,bib:HESS,bib:Magic,bib:Pamela,bib:AMS}. 

Gamma-ray telescopes provide very strong con\-straints on the thermally-averaged annihilation cross section from observations of satellite dwarf spheroidal galaxies~\cite{bib:Fermi_Magic_dwarfs}. However, neutrinos are the only messenger particles that can be used to probe for dark matter in close-by massive baryonic bodies like the Sun or the Earth. In these objects dark matter particles from the Galactic halo can be accumulated after becoming bound in the gravitational potential of the Solar system as it passes through the Galaxy~\cite{bib:Gould,bib:Freese,bib:Press,bib:Gaisser}. The WIMPs may then scatter weakly on nuclei in the celestial bodies and lose energy. Over time, this leads to an accumulation of dark matter in the center of the bodies. The accumulated dark matter may then self-annihilate at a rate that is proportional to the square of its density, generating a flux of neutrinos with a spectrum that depends  on the annihilation channel and WIMP mass. The annihilation would also contribute to the energy deposition in the Earth. A comparison of the expected energy deposition with the measured heat flow allows to exclude strongly interacting dark matter~\cite{bib:head}.

The expected neutrino event rates and energies depend on the specific nature of dark matter, its local density and velocity distribution, and the chemical composition of the Earth. Different scenarios yield neutrino-induced muon fluxes between $10^{-8}-10^5$ per km$^2$ per year for WIMPs with masses in the GeV$-$TeV range~\cite{bib:Bruch}. The AMANDA~\cite{bib:AMANDAsearch,bib:amandaanalysis} and Super-K~\cite{bib:SuperK} collaborations have already ruled out muon fluxes above $\sim10^3$ per km$^2$ per year for masses larger than some 100 GeV. The ANTA\-RES collaboration has recently presented the results of a similar search using five years of data~\cite{bib:ANTARESsearch}. The possibility of looking for even smaller fluxes with the much bigger IceCube neutrino observatory motivates the continued search for neutrinos coming from WIMP annihilations in the center of the Earth. This search is sensitive to the spin-independent WIMP-nucleon cross section and complements IceCube searches for dark matter in the Sun~\cite{bib:IC79Solar}, the Galactic center~\cite{bib:IC79GC} and halo~\cite{bib:IC79Halo} and in dwarf spheroidal galaxies~\cite{bib:IC59dwarfs}.

\section{The IceCube Neutrino Telescope \label{sec:IceCube}}
The IceCube telescope, situated  at the geographic South Pole, is designed to detect the Cherenkov radiation produced by high energy neutrino-induced charged leptons traveling through the detector volume. By recording the number of Cherenkov photons and their arrival times, the direction and energy of the charged lepton, and consequently that of the parent neutrino, can be reconstructed. 

IceCube consists of approximately 1~km$^3$ volume of ice instrumented with 5160 digital optical modules (DOMs)~\cite{bib:DOM} in 86~strings, deployed between 1450~m and 2450~m depth~\cite{bib:IceCube}. Each DOM contains a 25.3 cm diameter Hamamatsu R7081-02 photomultiplier tube~\cite{bib:PMT} connected to a waveform recording data acquisition circuit. The inner strings at the center of IceCube comprise DeepCore~\cite{bib:DeepCore}, a more densely instrumented sub-array equipped with higher quantum efficiency DOMs.

While the large ice overburden above the detector provides a shield against downward going, cosmic ray induced muons with energies~$\lesssim$~500~GeV at the surface, most analyses focus on upward going neutrinos employing the entire Earth as a filter. Additionally, low energy analyses use DeepCore as the fiducial volume and the surrounding IceCube strings as an active veto to reduce penetrating muon backgrounds. The search for WIMP annihilation signatures at the center of the Earth takes advantage of these two background rejection techniques as the expected signal will be vertically up-going and of low energy. 

\section{Neutrinos from dark matter annihilations in the center of the Earth}
WIMPs annihilating in the center of the Earth will produce a unique signature in IceCube as vertically up-going muons. The number of detected neutrino-induced muons depends on the WIMP annihilation rate $\Gamma_A$. If the capture rate $C$ is constant in time $t$, $\Gamma_A$ is given by~\cite{bib:Bruch}

\begin{equation}
\Gamma_A = \frac{C}{2} \tanh^2\left(\frac{t}{\tau}\right), \ \tau = \left(CC_A\right)^{-1/2}.
\end{equation}

The equilibrium time $\tau$ is defined as the time when the annihilation rate and the capture rate are equal. $C_A$ is a constant depending on the WIMP number density. For the Earth, the equilibrium time is of the order of $10^{11}$~years if the spin-independent WIMP-nucleon cross section is $\sigma_{\chi-N}^\text{SI} \sim 10^{-43} ~ \rm{cm}^2$~\cite{bib:JKG}. The age of the Solar system is $t_\circ \approx 4.5\cdot 10^{9}$~years and so $t_\circ/\tau \ll 1$. We thus expect that $\Gamma_A \propto C^2$, i.e. the higher the capture rate, the higher the annihilation rate and thus the neutrino-induced muon flux.

\begin{figure}[!htb]
  \centering
  \includegraphics[width=0.45\textwidth]{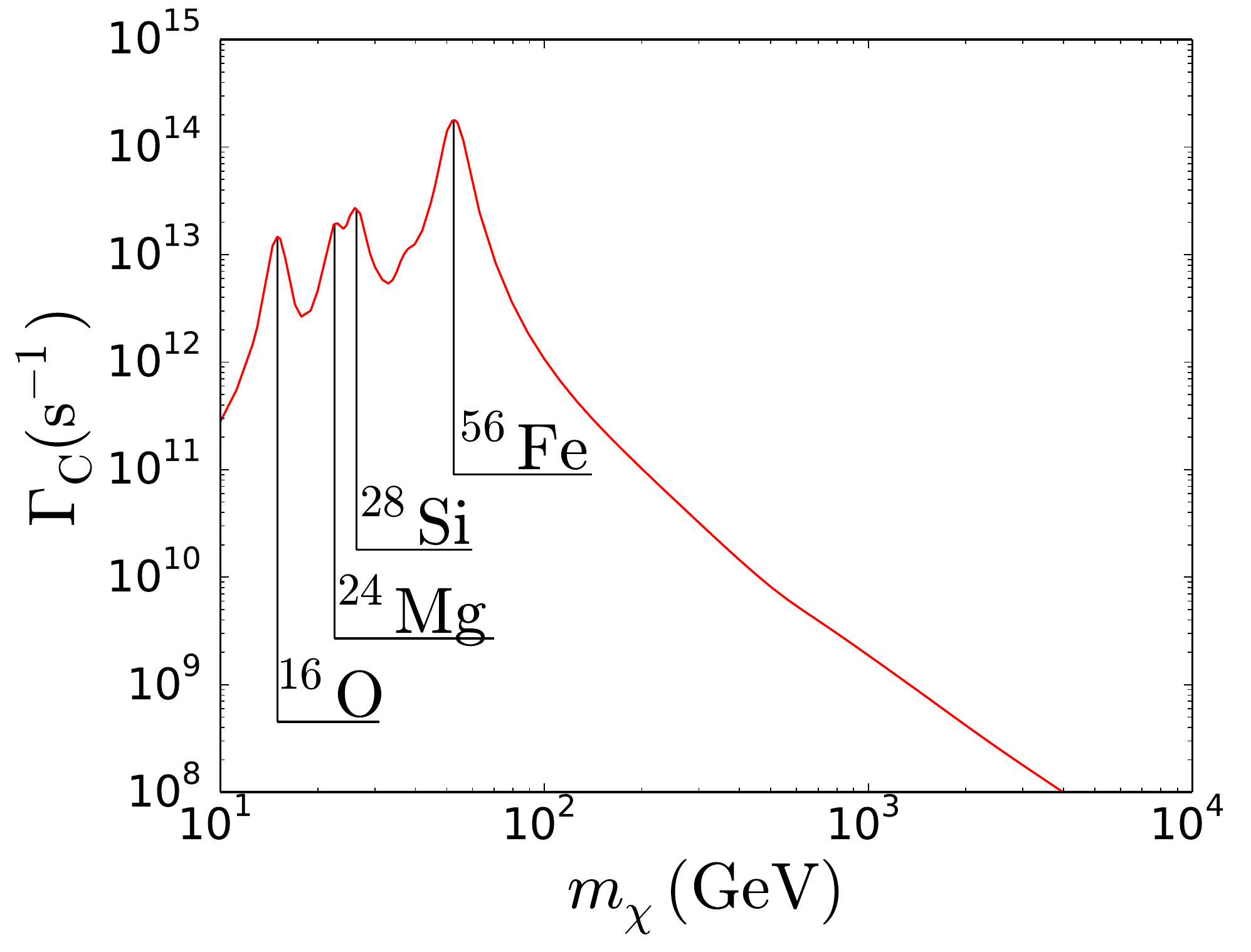}
  \caption{Rate at which dark matter particles are captured to the interior of the Earth~\cite{bib:edsjo2} for a scattering cross section of $\sigma_\text{SI} = 10^{-44}$~cm$^2$ . The peaks correspond to resonant capture on the most abundant elements in the Earth~\cite{bib:earth-model}:  $^{56}$Fe, $^{16}$O, $^{28}$Si and $^{24}$Mg and their isotopes.}
  \label{fig:capture_rate}
\end{figure}

The rate at which WIMPs are captured in the Earth depends on their mass (which is unknown), their velocity in the halo (which can not be measured observationally, and therefore needs to be estimated through simulations) and their local density (which can be estimated from observations). The exact value of the local dark matter density is still under debate~\cite{bib:read_density}, with estimations ranging from $\sim 0.2~{\rm GeV}/{\rm cm}^3$ to $\sim0.5~{\rm GeV}/\rm{cm}^3$. We take a value of 0.3~GeV/cm$^3$ as suggested in~\cite{bib:PDG} for the results presented in this paper in order to compare to the results of other experiments. If the WIMP mass is nearly identical to that of one of the nuclear species in the Earth,
the capture rate will increase considerably, as shown in Fig.~\ref{fig:capture_rate}.

The capture rate could be higher if the velocity distribution of WIMPs with respect to the Earth is lower, as only dark matter with lower velocities can be captured by the Earth. The velocity distribution of dark matter in the halo is uncertain, as it is very sensitive to theoretical assumptions. The simplest halo model is the Standard Halo Model (SHM), a smooth, spherically symmetric density component with a non-rotating Gaussian velocity distribution~\cite{bib:SHM}. Galaxy formation simulations indicate, however, that additional macro\-structural components, like a dark disc~\cite{bib:Lake,bib:Read1,bib:Read2}, could exist. This would affect the velocity distribution, especially at low velocities, and, consequently, the capture rate in the Earth.

The signal simulations that are used in the analysis are performed using WimpSim~\cite{bib:WimpSim}, which describes the capture and annihilation of WIMPs inside the Earth, collects all neutrinos that emerge and lets these propagate through the Earth to the detector. The code includes neutrino interactions and neutrino oscillations in a complete three-flavor treatment. Eleven benchmark masses between 10\,GeV and 10\,TeV were simulated for different annihilation channels: the annihilation into $b\bar{b}$ leads to a soft neutrino energy spectrum, while a hard channel is defined by the annihilation into $W^+W^-$ for WIMP masses larger than the rest mass of the $W$ bosons and annihilation into $\tau^+\tau^-$ for lower WIMP masses. 

\section{Background}
As signal neutrinos originate near the center of the Earth, they induce a vertically up-going signal in the detector. This is however a special direction in the geometry of IceCube, as the strings are also vertical. While in other point source searches, a signal-free control region of the same detector acceptance can be defined by changing the azimuth, this is not possible for an Earth WIMP analysis. Consequently, a reliable background estimate can only be derived from simulation.

Two types of background have to be taken into account: the first type consists of atmospheric muons produced by cosmic rays in the atmosphere above the detector. Although these particles enter the detector from above, a small fraction will be reconstructed incorrectly as up-going. The cosmic ray interactions in the atmosphere that produce these particles are simulated by CORSIKA~\cite{bib:corsika}. 

The second type of background consists of atmospheric neutrinos. This irreducible background is coming from all directions and is  simulated with GENIE~\cite{bib:genie} for neutrinos with energies below 190~GeV and with NuGeN~\cite{bib:nugen} for higher energies. 

\section{Event selection}
\label{sec:event_selection}
This analysis used the data taken in the first year of the fully deployed detector (from May 2011 to May 2012) with a livetime of 327 days. During the optimization of the event selection, only 10\% of the complete dataset was used to check the agreement with the simulations. The size of this dataset is small enough to not reveal any potential signal, and hence allows us to maintain statistical blindness.

 To be sensitive to a wide range of WIMP masses, the analysis is split into two parts that are optimized separately. The high energy event selection aims for an optimal sensitivity for WIMP masses of 1\,TeV and the $\chi\chi\rightarrow W^+W^-$ channel. The event selection for the low energy part is optimized for 50~GeV WIMPs annihilating into tau leptons. Because the capture rate for WIMPs of this mass shows a maximum (see Fig.~\ref{fig:capture_rate}), the annihilation and thus the expected neutrino rate are also maximal. As the expected neutrino energy for 50\,GeV WIMPs is lower than 50\,GeV, the DeepCore detector is crucial in this part of the analysis. Both samples are analyzed for the hard and the soft channel. 

The data are dominated by atmospheric muons (kHz rate),  which can be reduced via selection cuts, as explained below. These cuts lower the data rate by six orders of magnitude, to reach the level where the data are mainly consisting of atmospheric neutrino events (mHz rate). Since atmospheric neutrino events are indistinguishable from signal if they have the same direction and energy as signal neutrino events, a statistical analysis is performed on the final neutrino sample, to look for an excess coming from the center of the Earth (zenith=$180^\circ$).

The first set of selection criteria, based on initial track reconstructions~\cite{bib:Reconstruction}, is applied on the whole data\-set, i.e.~before splitting it into a low and a high energy sample. This reduces the data rate to a few Hz, so that more precise (and more time-consuming) reconstructions can be used to calculate the energy on which the splitting will be based. These initial cuts consist of a selection of online filters that tag up-going events, followed by cuts on the location of the interaction vertex and the direction of the charged lepton. These variables are not correlated with the energy of the neutrino and have thus similar efficiencies for different WIMP masses.

The variables that are used for cuts at this level are the reconstructed zenith angle, the reconstructed interaction vertex and the average temporal development of hits in the vertical ($z$) direction. The zenith angle cut is relatively loose to retain a sufficiently large control region in which the agreement between data and background simulation can be tested. An event is removed if the reconstructed direction points more than 60$^\circ$ from the center of the Earth (i.e. the zenith is required to be larger than $120^\circ$). In this way the agreement between data and background simulation can be tested in a signal-free zenith region between 120$^\circ$ and 150$^\circ$ (see zenith distribution in Fig.~\ref{fig:zenith_distributions}). The other cut values are chosen by looping over all possible combinations and checking which combination brings down the background to the Hz level, while removing as little signal as possible. 

\begin{figure}[tb]
  \centering
  \includegraphics[width=0.5\textwidth]{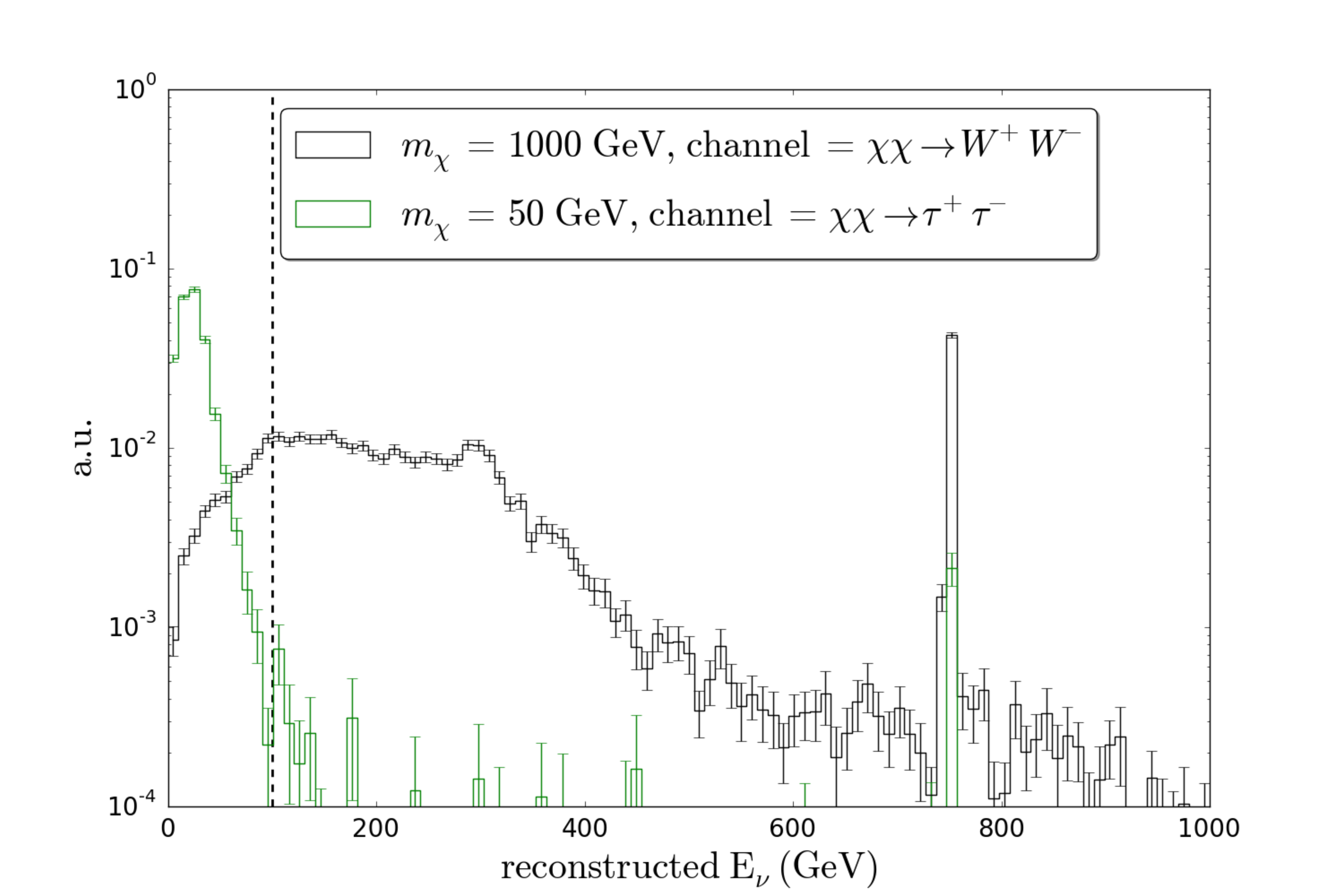}
  \caption{Reconstructed energy distributions for neutrinos induced by 50~GeV and 1~TeV WIMPs trapped in the Earth. The vertical dashed line shows where the dataset is split. The error bars show statistical uncertainties. See Sec.\ref{sec:event_selection} for an explanation of the peak at 750~GeV.}
  \label{fig:energy_reco}
\end{figure}

After this first cut level, the data rate is reduced to $\sim$3~Hz, while 30\%-60\% of the signal (depending on WIMP mass and channel) is kept. The data is still dominated by atmospheric muons at this level. Now that the rate is sufficiently low, additional reconstructions can be applied to the data~\cite{bib:LEERA}. 

The distribution of the reconstructed energies for 50 GeV and 1 TeV WIMP signal events are shown in Fig.~\ref{fig:energy_reco}.
The peak at $\sim$750~GeV is an artifact of the energy reconstruction algorithm used in this analysis: if the track is not contained in the detector, the track length cannot be reconstructed and is set to a default value of 2~km. The track length is used to estimate the energy of the produced muon, while the energy of the hadronic cascade is reconstructed separately and can exceed the muon energy. Events showing this artifact are generally bright events, so their classification into the high energy sample is desired. The reconstructed energy is not used for other purposes than for splitting the data. A division at 100 GeV, shown as a vertical line in this figure, is used to split the dataset into low and high energy samples which are statistically independent and are optimized and analyzed separately.

\begin{figure*}[tb!]
  \centering
  \includegraphics[width=1\columnwidth]{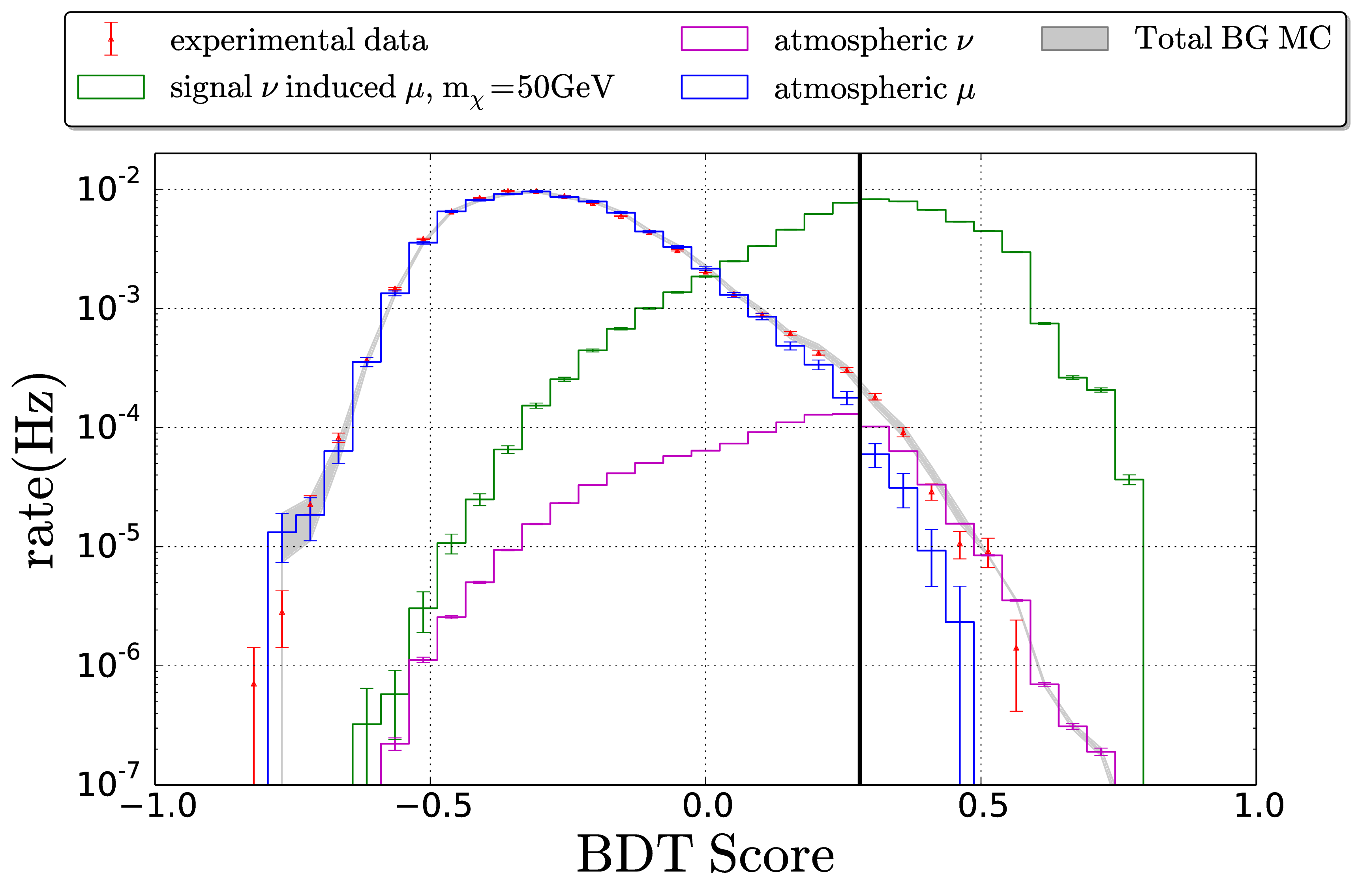}
  \includegraphics[width=1\columnwidth]{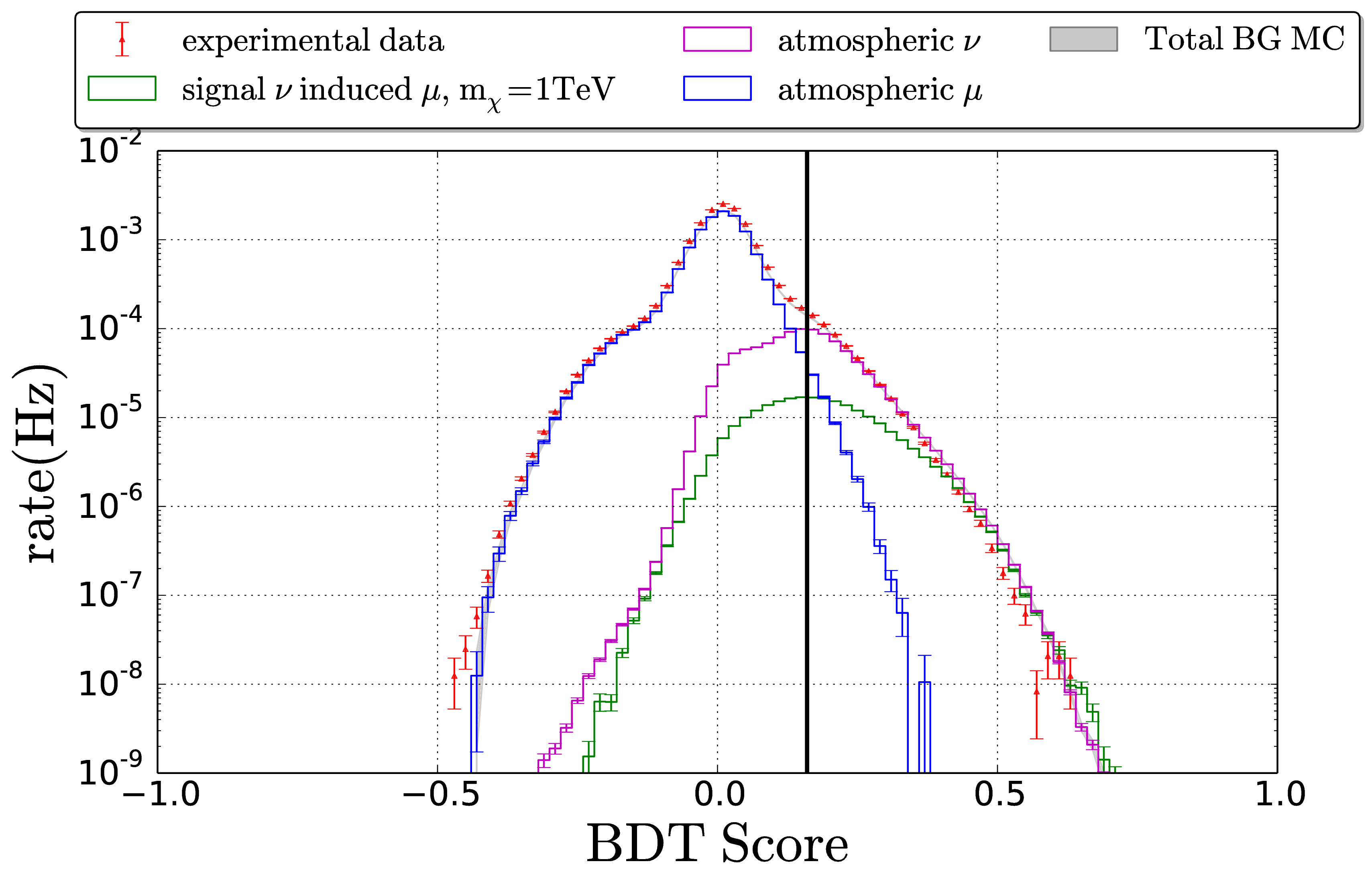}
  \caption{BDT score distributions at pre-BDT level for the low energy analysis (left) and for the high energy analysis using the {\it Pull-Validation} method (right).  Signal distributions are upscaled to be visible in the plot.  Signal and backgrounds are compared to experimental data from 10\% of the first year of IC86 data. For the atmospheric neutrinos, all flavors are taken into account. In gray, the sum of all simulated background is shown. The vertical lines indicate the final cut value used in each analysis, where high scores to the right of the line are retained.}
  \label{fig:BDT_distributions}
\end{figure*}
Both analyses use Boosted Decision Trees (BDTs) to classify background and signal events. This machine learning technique is designed to optimally separate signal from background after an analysis-specific training~\cite{bib:TMVA} by assigning a score between -1 (background-like) and +1 (signal-like) to each event. In order to train a reliable BDT, the simulation must reproduce the experimental data accurately. Therefore a set of pre-BDT cuts are performed. Demanding a minimum of hits in a time window between -15 ns and 125 ns of the expected photon arrival time at each DOM, and
a cut on the zenith of a more accurate reconstruction on causally connected hits improves the agreement between data and simulation. By comparing the times and distances of the first hits, the number of events with noise hits can be reduced. The last cut variable at this step is calculated by summing the signs of the differences between the $z$-coordinates of two temporally succeeding hits, which reduces further the amount of misreconstructed events. After these cuts, the experimental data rates are of the order of 100~mHz, and the data are still dominated by atmospheric muons. The BDTs are then trained on variables that show good agreement between data and simulation and have low correlation between themselves.

In the low energy optimization, the BDT training samples consist of simulated 50~GeV WIMP events and experimental data for the signal and background respectively. Because the opening angle between the neutrino and its daughter lepton is inversely correlated to the energy of the neutrino, WIMP neutrino-induced muons in the high energy analysis are narrowly concentrated into vertical zenith angles, whereas in the low energy analysis they are spread over a wider range of zenith angles. Consequently, if the BDT for the high energy optimization was trained on simulated 1~TeV WIMP events, straight vertical events would be selected. This would make a comparison between data and simulation in a signal-free region more difficult. Instead, in the high energy analysis an isotropic muon neutrino simulation weighted to the energy spectrum of 1~TeV signal neutrinos is used to train a BDT.

Coincident events of neutrinos and atmospheric mu\-ons can affect the data rate. Their influence is larger at low energies, as the atmospheric neutrino flux decreases steeply with increasing energy. In the low energy analysis, this effect cannot be neglected. As the amount of available simulated coincident events was limited, individual correction factors for the components of atmospheric background simulation are applied to take this effect into account. These correction factors are calculated by scaling the BDT score distributions of the simulated background to the experimental data. Only events with a reconstructed zenith of less than 132$^\circ$ are used to determine the correction factors.  With this choice, the background cannot be incorrectly adjusted to a signal that could be contained in the experimental data, as  95\% of WIMP induced events have a larger zenith.

The distributions of the BDT scores for the low energy and high energy analyses are shown in Fig.~\ref{fig:BDT_distributions}. Cuts on the BDT score are chosen such that the sensitivities of the analyses are optimal. The sensitivities are calculated with a likelihood ratio hypothesis test based on the values of the reconstructed zenith, using the Feldman-Cousins unified approach~\cite{bib:FC}. The required probability densities for signal and background are both calculated from simulations, as this analysis cannot make use of an off-source region. The background sample that is left after the cut on the BDT score mainly consists of atmospheric neutrinos and only has a small number of atmospheric muon events. 

Due to small statistics of simulation we found it necessary to apply the smoothing techniques described in the following. The high energy analysis uses {\it Pull-Validation}~\cite{bib:pullvalidation}, a method to improve the usage of limited statistics: A large number of BDTs (200 in the case of the present analysis) are trained on small subsets that are randomly resampled from the complete dataset. The variation of the BDT output between the trainings can be interpreted as a probability density function (PDF) for each event. This PDF can be used to calculate a weight that is applied to each event instead of making a binary cut decision. With this method, not only the BDT score distribution is smoothed (Fig.~\ref{fig:BDT_distributions}-right), but also the distributions that are made after a cut on the BDT score. In particular, the reconstructed zenith distribution used in the likelihood calculation is smooth, as events that would be removed when using a single BDT could now be kept, albeit with a smaller weight.

The low energy analysis tackles the problem of poor statistics of the atmospheric muon background simulation in a different way. In this part of the analysis, only a single BDT is trained (Fig.~\ref{fig:BDT_distributions}-left), and after the cut on the BDT score, the reconstructed zenith distribution is smoothed using a Kernel Density Estimator (KDE)~\cite{bib:KDE_Rosenblatt,bib:KDE_Parzen} with gaussian kernel and choosing an optimal bandwidth~\cite{bib:Silverman}.

The event rates at different cut levels are summarized in Table~\ref{tab:rates}.

\begin{table*}[!htb]
  \centering
  \begin{tabular}{ c | c | c | c | c | c | c | c | c }
    \hline
    Cut        & \multicolumn{2}{|c}{Data}          & \multicolumn{2}{|c}{Atm. $\mu$}        & \multicolumn{2}{|c}{Atm. $\nu$}        & \multicolumn{2}{|c}{Signal}          \\
    level      & \multicolumn{2}{|c}{rate [Hz]}     & \multicolumn{2}{|c}{rate [Hz]}         & \multicolumn{2}{|c}{rate [Hz]}         & \multicolumn{2}{|c}{eff.}       \\
    \hline
      & LE & HE & LE & HE & LE & HE & LE & HE\\
    \hline
    2 & \multicolumn{2}{|c}{670} &\multicolumn{2}{|c}{650} & \multicolumn{2}{|c}{0.027} & \multicolumn{2}{|c}{100\,\%} \\
    3 & 1.39 & 1.35 & 1.03 & 0.97 & 2.5$\cdot 10^{-3}$ & 2.0$\cdot 10^{-3}$ & 40.8\,\% & 45.1\,\% \\
    4 & 2.8$\cdot 10^{-4}$ & 5.6$\cdot 10^{-4}$  & 8.0$\cdot 10^{-5}$ & 6.3$\cdot 10^{-5}$ & 2.0$\cdot 10^{-4}$ & 4.6$\cdot 10^{-4}$ & 15.6\,\% & 17.0\,\% \\
    \hline
  \end{tabular}
  \caption{Rates for experimental data, simulated atmospheric muons and atmospheric neutrinos of all favors, and signal efficiencies for WIMP masses of 50 GeV and 1 TeV, respectively, at different cut levels. Level 2 refers to the predefined common starting level, level 3 shows the event rates after the first set of cuts and the split into a high (HE) and a low energy (LE) sample and level 4 indicates the final analysis level after additional cuts and the BDT selection. Note that due to the Pull-Validation procedure, all events in the high energy sample at final level contain a weight. The effective data rates are shown.}
  \label{tab:rates}
\end{table*}

\section{Shape analysis}
\label{sec:method}
After the event selection, the data rate is reduced to 0.28\,mHz for the low energy selection and 0.56\,mHz for the high energy selection. Misreconstructed atmospheric muons are almost completely filtered out and the remaining data sample consists mainly of atmospheric neutrinos. To analyze the dataset for an additional neutrino signal coming from the center of the Earth, we define a likelihood test, that has been used in several IceCube analyses before (e.g.~\cite{bib:IC79Solar,bib:IC79GC}). Based on the background ($f_{bg}$) and signal distribution($f_{s}$) of space angles $\Psi$ between the reconstructed muon track and the Earth center (i.e. the reconstructed zenith angle), 
the probability to observe a value $\Psi$ for a single event is
\begin{equation}
  f(\Psi|\mu)=\frac{\mu}{n_\text{obs}}f_s(\Psi)+(1-\frac{\mu}{n_\text{obs}})f_{bg}(\Psi)\,.
\end{equation}
Here, $\mu$ specifies the number of signal events in a set of $n_\text{obs}$ observed events. 
The likelihood to observe a certain number of events at specific space angles $\Psi_i$ is defined as
\begin{equation}
  \mathcal{L}=\prod_{i}^{n_\text{obs}}f(\Psi_i|\mu)\, .
\end{equation}
Following the procedure in~\cite{bib:FC}, the ranking parameter
\begin{equation}
  \mathcal{R}(\mu)=\frac{\mathcal{L}(\mu)}{\mathcal{L}(\hat{\mu})}
\end{equation}
is used as test statistic  for the hypothesis testing, where $\hat{\mu}$ is the best fit of $\mu$ to the observation.
A critical ranking $\mathcal{R}^{90}$ is defined for each signal strength, so that 90\% of all experiments have a ranking larger than $\mathcal{R}^{90}$. This is determined by $10^4$ pseudo experiments for each injected signal strength. The sensitivity is defined as the expectation value for the upper limit in case that no signal is present. This is determined by generating $10^4$ pseudo experiments with no signal injected.

\section{Systematic uncertainties}
\begin{figure}[!htb]
  \centering
  \includegraphics[width=0.4\textwidth]{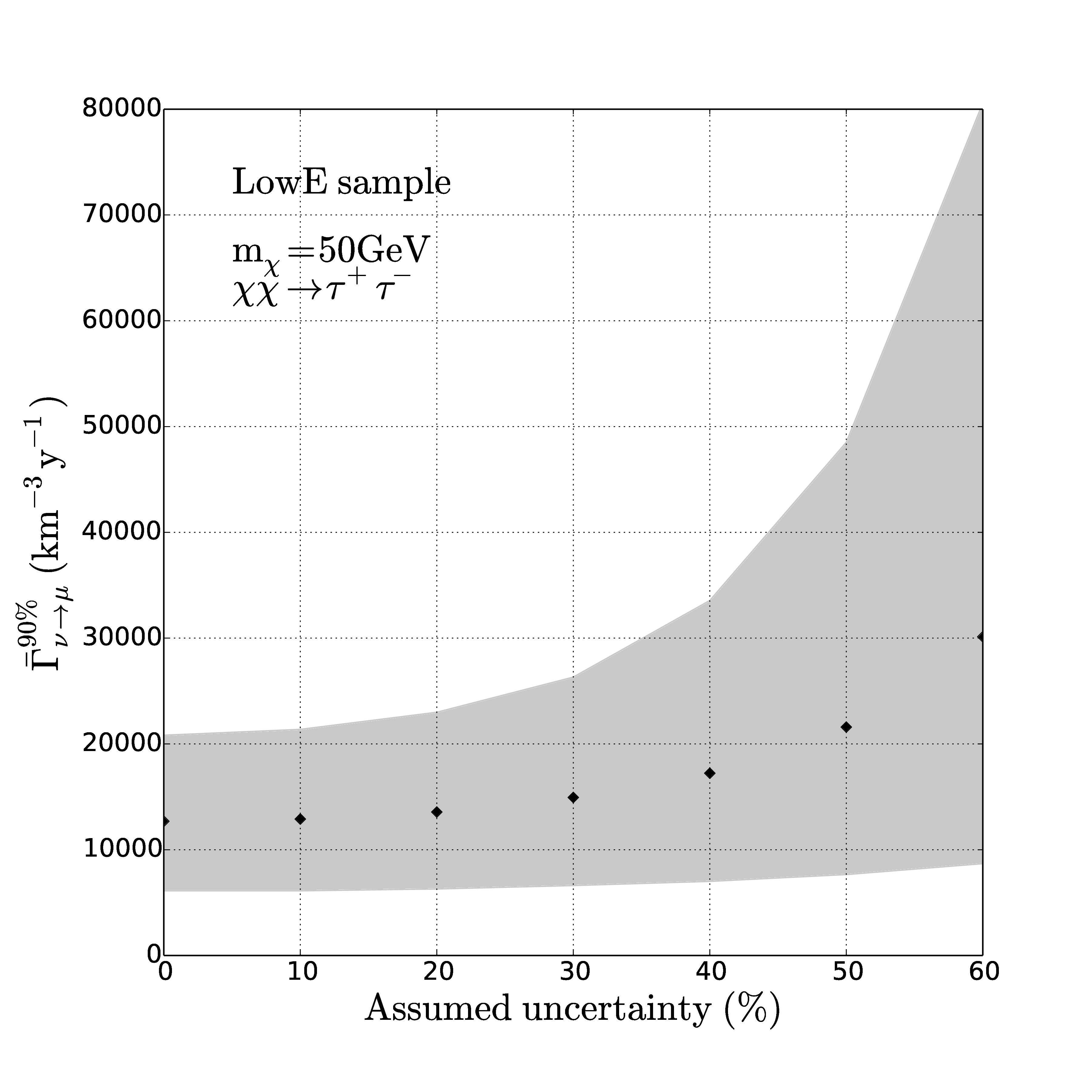}
  \caption{Effect of the assumed uncertainty on the sensitivity of the volumetric flux. The example shows 50\,GeV WIMPs annihilating into $\tau^+\tau^-$. The points show the estimated sensitivity and include a correction for coincident muons, while the band indicates one standard deviation.}
  \label{fig:uncertainties}
\end{figure}

Due to the lack of a control region, the background estimation has to be derived from simulation. Therefore, systematic uncertainties of the simulated datasets were carefully studied. The effects of the uncertainties were quantified by varying the respective input parameters in the simulations.

Different types of detector related uncertainties have to be considered. The efficiency of the DOM to detect Cherenkov photons is not exactly known. To estimate the effect of this uncertainty, three simulated data sets with 90\%, 100\% and 110\% of the nominal efficiency were investigated. With these data sets, the sensitivity varies by $\pm$10\% for both event selections of the analysis. Taking anisotropic scattering in the South Pole ice into account~\cite{bib:SpiceLea}, has an effect of $-$10\% in the high and the low energy selection. The reduced scattering length of photons in the refrozen ice of the holes leads to an uncertainty of $-$10\% in both selections. Furthermore, the uncertainty on the scattering and absorption lengths influences the result by $\pm$10\% for the low energy and $\pm$5\% for the high energy selection. 

Besides the detector related uncertainties, the uncertainties on the models of the background physics are taken into account. The uncertainty of the atmospheric flux can change the rates by $\pm$30\%, as determined e.g. in~\cite{bib:ICcosmicRays}. For low energies, uncertainties on neutrino oscillation parameters are significant. This effect has been studied in a previous analysis~\cite{bib:IC79Solar} and influences the event rates by $\pm$6\%. The effect of the uncertainty of the neutrino-nucleon cross section has been studied in the same analysis. It depends on the neutrino energy and is conservatively estimated as $\pm$6\% for the low and $\pm$3\% for the high energy sample. Finally, the rate of coincidences of atmospheric neutrinos and atmospheric muons has a large impact on the low energy analysis. While in the baseline data sets, coincident events were not simulated, a comparison with a test simulation that includes coincident events shows an effect of $-$30\% on the final event rates.

Adding these uncertainties in quadrature results in a total of +34\%/-48\% in the low energy analysis and +32\%/-35\% for high energies. For the limit calculation, they are taken into account by using a semi-bayesian extension to the Feldman-Cousins approach~\cite{bib:conrad}. Technically, it is realized by randomly varying the expectation value of each pseudo-experiment by a gaussian of the corresponding uncertainty. As an illustration, the effect of this procedure is shown in Fig.~\ref{fig:uncertainties} for different uncertainties. 

\section{Results}
\label{sec:results}
\begin{figure*}[hbt!]
  \centering
  \includegraphics[width=0.91\columnwidth]{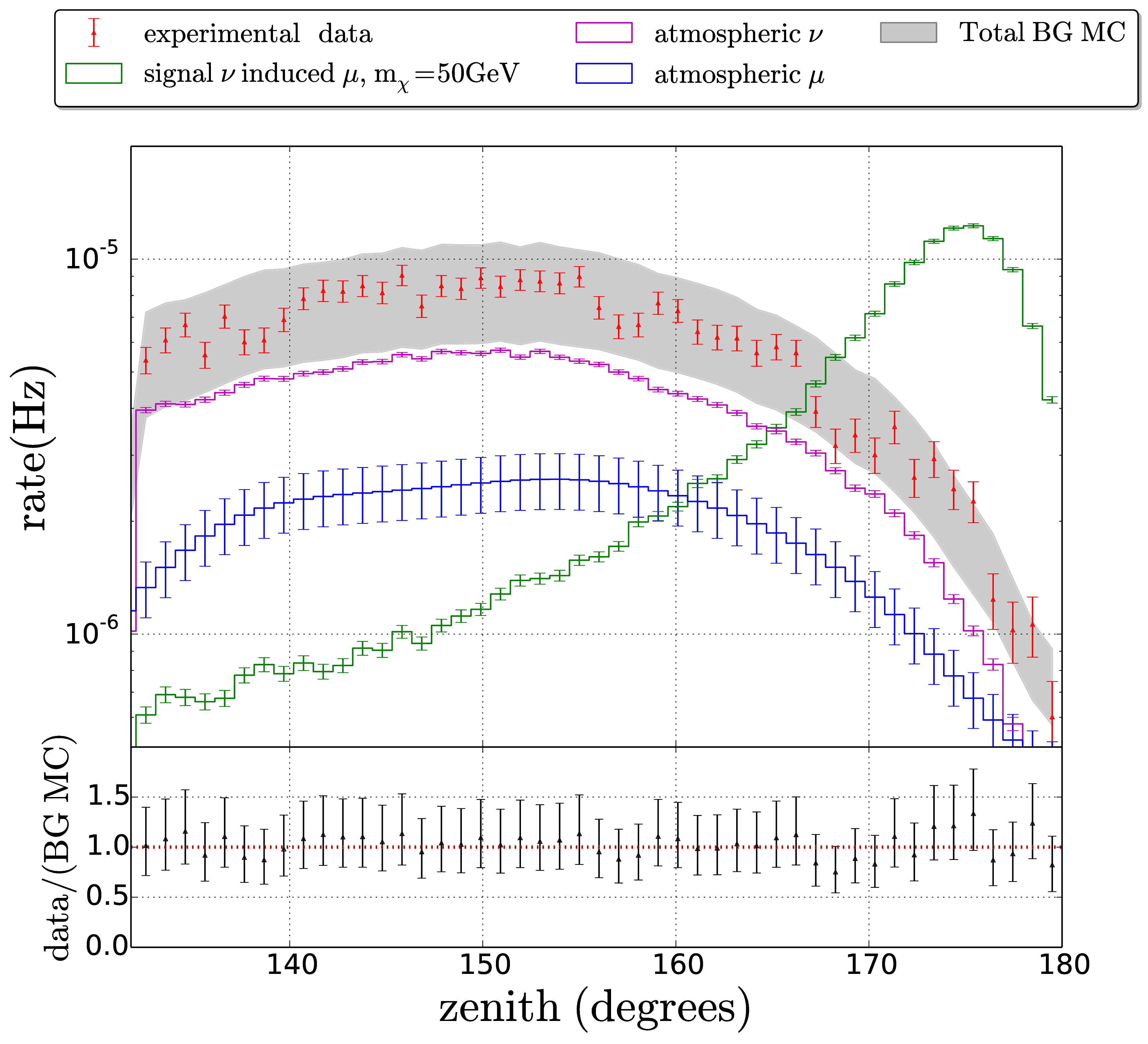}
  \includegraphics[width=0.91\columnwidth]{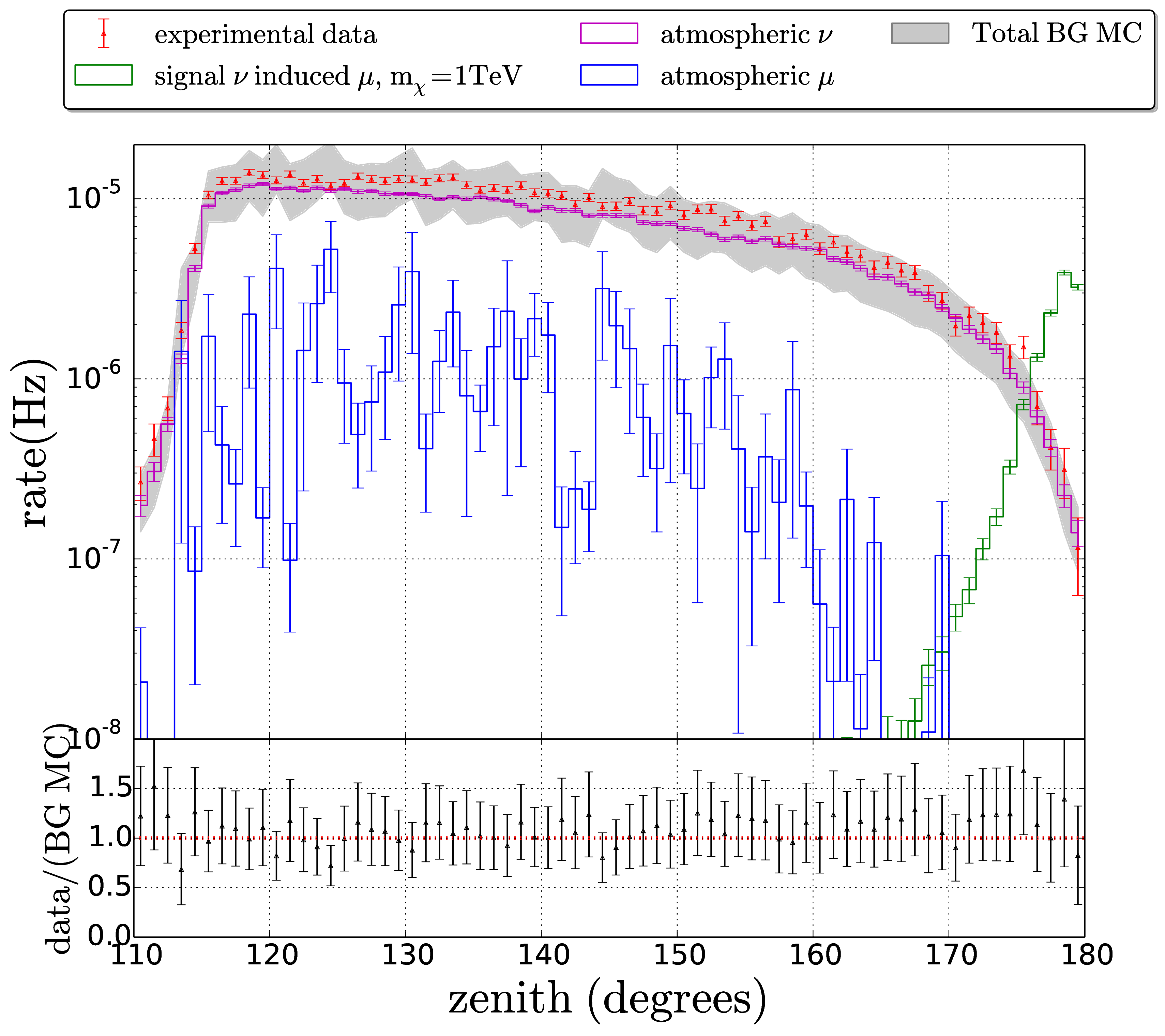}
  \caption{Reconstructed zenith distributions of 1 year of IC86 data (statistical uncertainties only) compared to the simulated background distributions, which include statistical and systematic uncertainties. For the atmospheric neutrinos, all flavors are taken into account. In the low energy analysis (left) the distributions were smoothed by a KDE and in the high energy analysis (right) the Pull-Validation method was used.  Signal distributions are upscaled to be visible in the plot. The gray areas indicate the total predicted background distributions with 1 sigma uncertainties, including statistical and systematic uncertainties.}
  \label{fig:zenith_distributions}
\end{figure*}
\begin{table*}[hbt!]
  \centering
  \begin{tabular}{ c | c | c | c | c | c | c | c}
    \hline 
    WIMP mass   & \multicolumn{2}{|c}{$\mu_s$} & \multicolumn{2}{|c}{$\Gamma_A$} & \multicolumn{2}{|c|}{$\Phi_\mu$} & $\sigma_\text{SI}$\\
    (GeV/c$^2$) & \multicolumn{2}{|c}{(year$^{-1}$)} & \multicolumn{2}{|c|}{(s$^{-1}$)}            & \multicolumn{2}{|c|}{(km$^{-2}$ year$^{-1}$)} & (cm$^{2}$) \\
    \hline 
         & hard channel & soft channel & hard channel & soft channel & hard channel & soft channel & hard channel\\
    \hline
    10    & 586 & -- & $3.01\cdot 10^{16}$ &       --           & $1.54\cdot 10^{4}$ &      --           & $2.5\cdot 10^{-38}$ \\ 
    20    & 209 & -- & $0.90\cdot 10^{15}$ &       --           & $3.57\cdot 10^{3}$ &      --           & $6.0\cdot 10^{-41}$ \\
    35    & 202 & 405 & $2.35\cdot 10^{14}$ & $4.05\cdot 10^{16}$ & $2.52\cdot 10^{3}$ & $8.70\cdot 10^{3}$ & $1.1\cdot 10^{-41}$  \\
    50    & 189 & 253 & $1.12\cdot 10^{14}$ & $7.88\cdot 10^{15}$ & $1.62\cdot 10^{2}$ & $3.85\cdot 10^{3}$ & $2.8\cdot 10^{-43}$  \\
    100   & 148 & 172 & $3.25\cdot 10^{13}$ & $5.24\cdot 10^{14}$ & $8.12\cdot 10^{2}$ & $1.36\cdot 10^{3}$ & $1.0\cdot 10^{-41}$  \\
    250   & 14.9 & 128 & $9.06\cdot 10^{11}$ & $4.22\cdot 10^{13}$ & $1.51\cdot 10^{2}$ & $7.30\cdot 10^{2}$ & $1.3\cdot 10^{-41}$  \\
    500   & 11.9 & 11.8 & $1.40\cdot 10^{11}$ & $3.49\cdot 10^{12}$ & 87.6              & $2.14\cdot 10^{2}$ & $1.7\cdot 10^{-41}$ \\
    1000  & 9.3 & 10.6 & $3.25\cdot 10^{10}$ & $5.38\cdot 10^{11}$ & 71.6              & $1.05\cdot 10^{2}$ & $2.0\cdot 10^{-41}$ \\
    3000  & 7.1 & 8.1 & $4.68\cdot 10^{9}$  & $6.88\cdot 10^{10}$ & 65.0              & 66.6              & $3.0\cdot 10^{-41}$ \\
    5000  & 6.6 & 7.5 & $2.12\cdot 10^{9}$  & $3.28\cdot 10^{10}$ & 64.1              & 60.3              & $3.8\cdot 10^{-41}$ \\
    10000 & 5.8 & 6.8 & $8.06\cdot 10^{8}$  & $1.47\cdot 10^{10}$ & 64.7              & 57.6              & $5.1\cdot 10^{-41}$ \\
    \hline
  \end{tabular}
  \caption{Upper limits at 90\% confidence level on the number of signal events $\mu_s$, the WIMP annihilation rate inside the Earth $\Gamma_A$, the muon flux $\Phi_\mu$ and the spin-independent cross section $\sigma_\text{SI}$, assuming an annihilation cross section of $\langle \sigma_A v \rangle = 3\cdot 10^{-26}\mathrm{{\rm cm}^{3}{\rm s}^{-1}}$. \emph{Soft channel} refers to annihilation into $b\bar{b}$, while \emph{hard channel} is defined by annihilation into $W^+W^-$ for WIMP masses larger than the rest mass of the W bosons and annihilation into $\tau^+\tau^-$ for lower WIMP masses. Systematic errors are included.}
  \label{tab:limits}
\end{table*}
As mentioned in Section~\ref{sec:event_selection}, only 10\% of the data were used for quality checks during the optimization of the analysis chain. Half of this subsample was used to train the BDTs and therefore these events could not be used for the later analysis. After the selection criteria were completely finalized, the zenith distributions of the remaining 95\% of the dataset were examined (Fig.~\ref{fig:zenith_distributions}). No statistically significant excess above the expected atmospheric background was found from the direction of the center of the Earth. 

\begin{figure}[hbt!]
  \centering
  \includegraphics[width=.9\columnwidth]{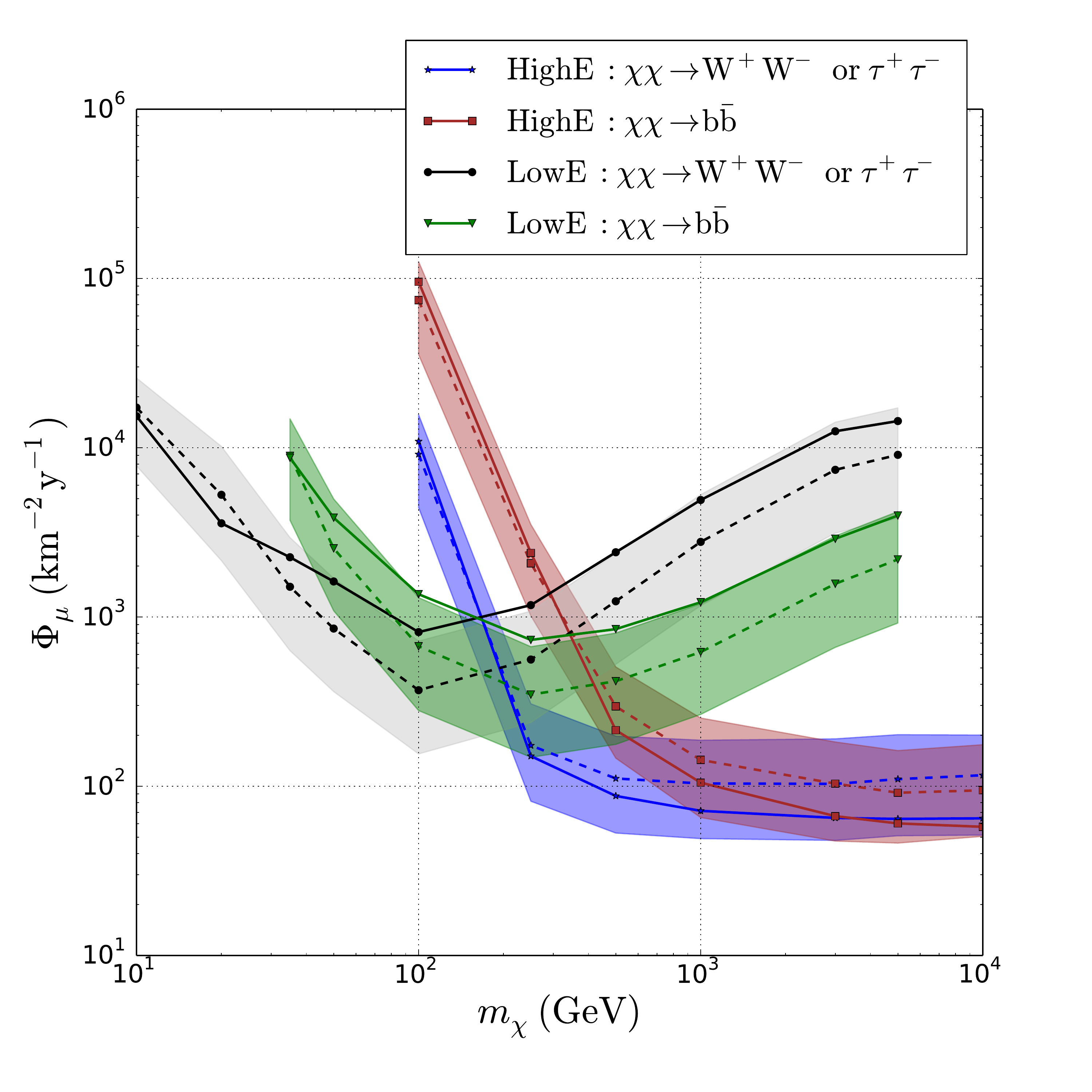}
  \includegraphics[width=.9\columnwidth]{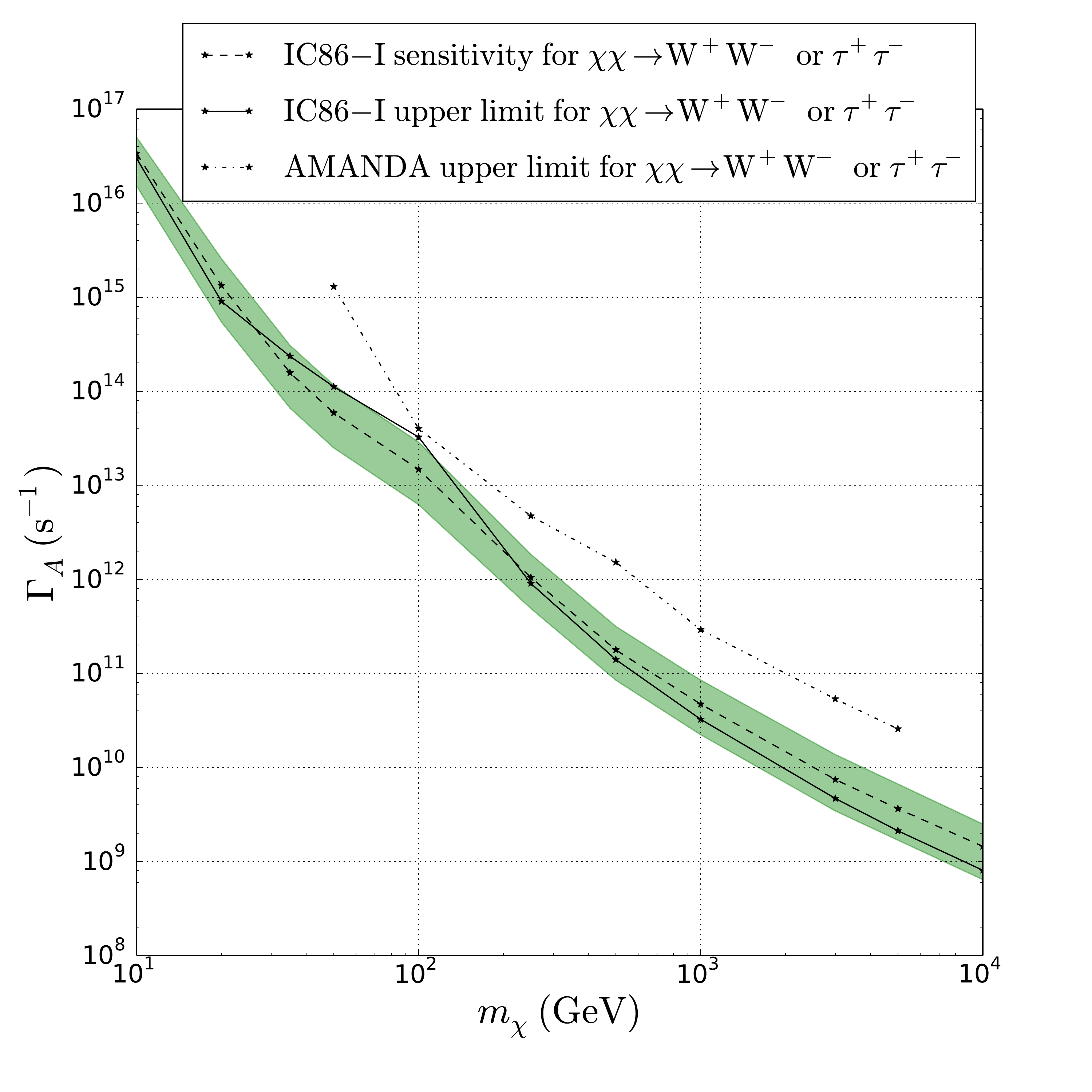}
  \caption{Top: Individual upper limits at 90\% confidence level (solid lines) on the muon flux $\Phi_\mu$ for the low and high energy analysis. Systematic uncertainties are included. For the soft channel, $\chi\chi\rightarrow b\bar{b}$ is assumed with 100\% braching ratio, while for the hard channel the annihilation $\chi\chi\rightarrow \tau^{+}\tau^{-}$ for masses $\leq$ 50 GeV and $\chi\chi\rightarrow W^{+}W^{-}$ for higher masses is assumed. A flux with mixed branching ratios will be between these extremes.
The dashed lines and the bands indicate the corresponding sensitivities with one sigma uncertainty. 
Bottom: The combined best upper limits (solid line) and sensitivities (dashed line) with 1 sigma uncertainty (green band) on the annihilation rate in the Earth $\Gamma_A$ for 1 year of IC86 data as a function of the WIMP mass. For each WIMP mass, the sample (high energy or low energy) which yields the best sensitivity is used. Systematic uncertainties are included. The dotted line shows the latest upper limit on the annihilation rate, which was calculated with AMANDA data~\cite{bib:AMANDAsearch,bib:amandaanalysis}.}
  \label{fig:limits}
\end{figure}

Using the method described in Section~\ref{sec:method}, upper limits at the 90\% confidence level on the volumetric flux 
\begin{equation}
  \Gamma_{\mu\rightarrow\nu}=\frac{\mu_s}{t_\mathrm{live}\cdot V_\mathrm{eff}}
\end{equation}
 were calculated from the high and the low energy sample for WIMP masses between 10\,GeV and 10\,TeV in the hard and in the soft channel. Here $\mu_s$ denotes the upper limit on the number of signal neutrinos, $t_\mathrm{live}$ the livetime and $V_\mathrm{eff}$ the effective volume of the detector. Using the package WimpSim~\cite{bib:WimpSim}, the volumetric flux was converted into the WIMP annihilation rate inside the Earth $\Gamma_A$ and the resulting muon flux $\Phi_\mu$. The obtained 90\% C.L. limits are shown in Fig.~\ref{fig:limits} and listed in Table~\ref{tab:limits}. For each mass and channel, the result with the most restricting limit is shown.

\begin{figure}[ht!]
  \centering
  \includegraphics[width=0.45\textwidth]{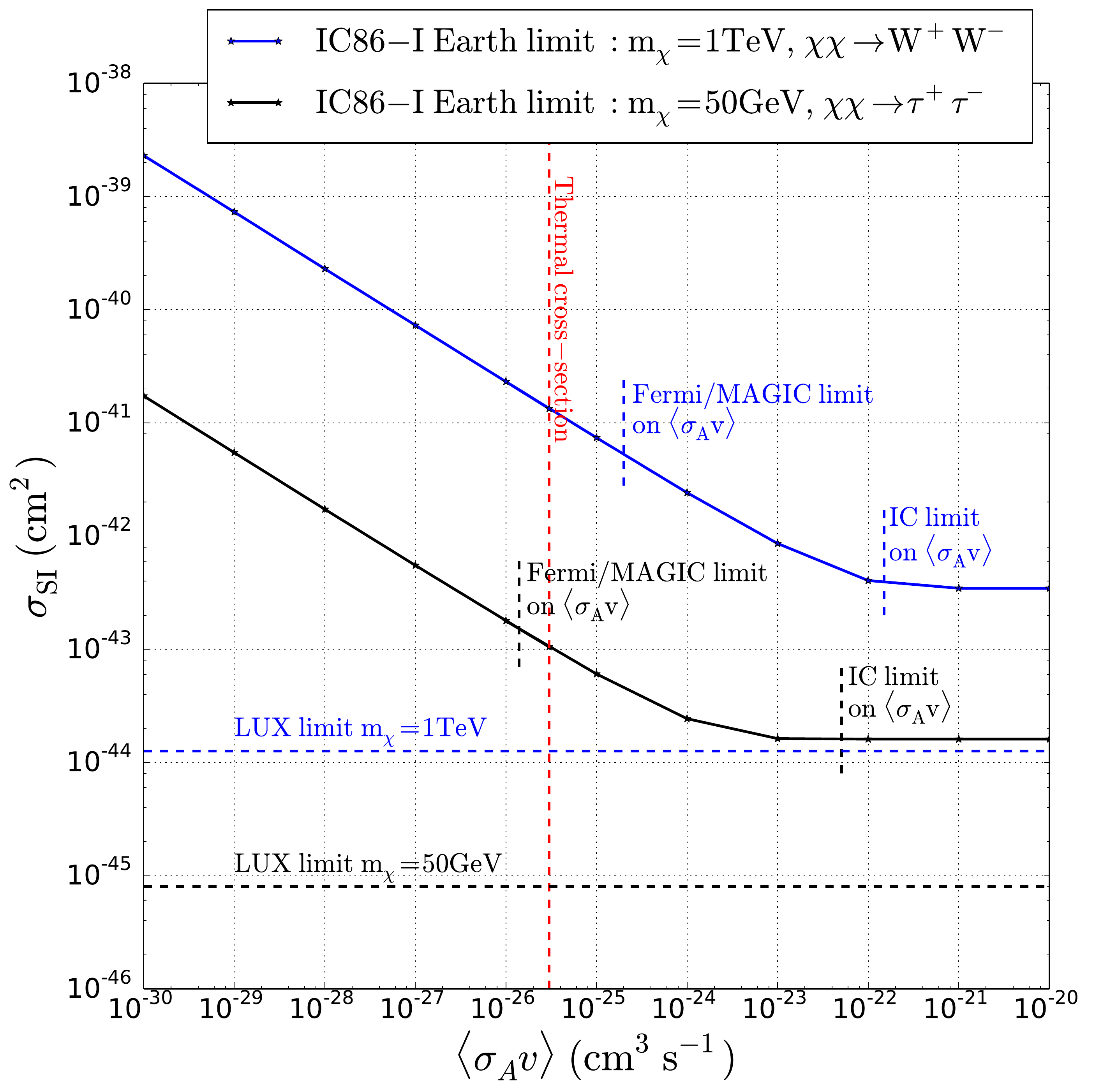}
  \caption{Upper limits at 90\% confidence level on $\sigma_{\chi-N}^\text{SI}$ as a function of the annihilation cross section for 50\,GeV WIMPs annihilating into $\tau^+\tau^-$ and for 1\,TeV WIMPs annihilating into $W^+W^-$. Systematic uncertainties are included. As a comparison, the limits of LUX~\cite{bib:LUX} are shown as dashed lines. The red vertical line indicates the thermal annihilation cross section. Also indicated are IceCube limits on the annihilation cross section for the respective models~\cite{bib:IC79GC}, as well as the limits from a combined analysis of Fermi-LAT and MAGIC~\cite{bib:Fermi_Magic_dwarfs}}
  \label{fig:sigma_si_vs_gammaA}
\end{figure}

\begin{figure}[hbt!]
  \centering
  \includegraphics[width=0.45\textwidth]{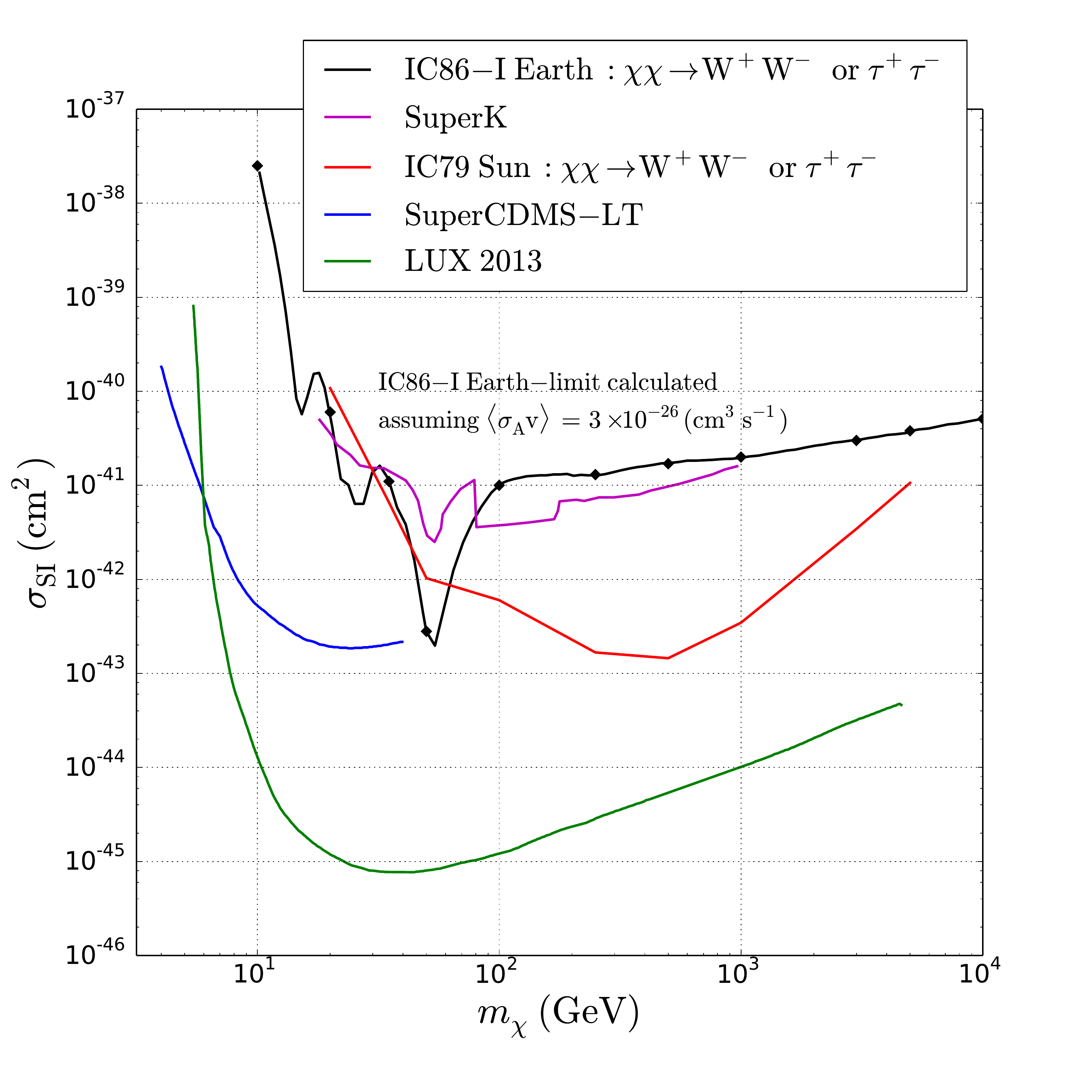}
  \caption{Upper limits at 90\% confidence level on  $\sigma_{\chi-N}^\text{SI}$ as a function of the WIMP-mass assuming a  WIMP annihilation cross section of  $\langle \sigma_A v \rangle = 3\cdot 10^{-26}\mathrm{{\rm cm}^{3}{\rm s}^{-1}}$. For WIMP masses above the rest mass of the $W$ bosons, annihilation into  $W^+W^-$ is assumed and annihilation into $\tau^+\tau^-$ for lower masses.  Systematic uncertainties are included. The result is compared to the limits set by SuperCDMSlite~\cite{bib:SuperCDMS}, LUX~\cite{bib:LUX}, Super-K~\cite{bib:SuperK} and by a Solar WIMP analysis of IceCube in the 79-string configuration~\cite{bib:IC79Solar}. The displayed limits are assuming a local dark matter density of  $\rho_\chi=0.3$~GeV~cm$^{-3}$. A larger density, as suggested e.g. by~\cite{bib:Nesti}, would scale all limits linearly.}
  \label{fig:sigma_si}
\end{figure}

Furthermore, limits on the spin-independent WIMP-nucleon cross section $\sigma_{\chi-N}^\text{SI}$ can be derived. In contrast to dark matter accumulated in the Sun, the annihilation rate in the Earth and $\sigma_{\chi-N}^\text{SI}$ are not directly linked. As no equilibrium between WIMP capture and annihilation can be assumed, the annihilation rate depends on $\sigma_{\chi-N}^\text{SI}$ and on the annihilation cross section $\langle \sigma_A v \rangle$. Fig.~\ref{fig:sigma_si_vs_gammaA} shows the limits in the $\sigma_{\chi-N}^\text{SI}$ - $\langle \sigma_A v \rangle$ plane for two WIMP masses. 
A typical value for the natural scale, for which the WIMP is a thermal relic~\cite{bib:relic_scale}, is $\langle \sigma_A v \rangle = 3\cdot 10^{-26}\,\mathrm{cm^{3}s^{-1}}$. 
To compute a limit on the spin-independent WIMP-nucleon scattering cross section that is consistent for all masses we use the thermal relic cross section, even though Fermi excludes this value at 95\%C.L for masses below about 80 GeV for $\tau\tau$ and $bb$ annihilation channels~\cite{bib:Fermi_Magic_dwarfs}.
While the limits in Table~\ref{tab:limits} correspond to the investigated benchmark masses, in Fig.~\ref{fig:sigma_si}, interpolated results were taken into account, showing the effect of the resonant capture on the most abundant elements in the Earth.

We note that Solar WIMP, Earth WIMP, and direct searches have very different dependences on astrophysical uncertainties. A change in the WIMP velocity distribution has minor effects on Solar WIMP bounds~\cite{bib:sun_boost1,bib:sun_boost2}, while Earth WIMPs and direct searches are far more susceptible to it. In particular the existence of a dark disk could enhance Earth WIMP rates by several orders of magnitude~\cite{bib:Bruch} while leaving direct bounds largely unchanged. The limits presented here assume a standard halo and are conservative with respect to the existence of a dark disk.

\section{Summary}
Using one year of data taken by the fully completed detector, we performed the first IceCube search for neutrinos produced by WIMP dark matter annihilations in the center of the Earth. No evidence for a signal was found and 90\% C.L. upper limits were set on the annihilation rate and the resulting muon flux as function of the WIMP mass.  Assuming the natural scale for the velocity averaged annihilation cross section, upper limits on the spin-independent WIMP-nucleon scattering cross section could be derived. The limits on the annihilation rate are up to a factor 10 more restricting than previous limits. For indirect WIMP searches through neutrinos, this analysis is highly complementary to Solar searches. In particular, at small WIMP masses around the iron resonance of 50\,GeV the sensitivity exceeds the sensitivity of the Solar WIMP searches of IceCube. The corresponding limit on the spin-in\-de\-pend\-ent cross sections presented in this paper are the best set by IceCube at this time. Future analyses combining several years of data will further improve the sensitivity.

\begin{acknowledgements}
We acknowledge the support from the following agencies:
U.S. National Science Foundation-Office of Polar Programs,
U.S. National Science Foundation-Physics Division,
University of Wisconsin Alumni Research Foundation,
the Grid Laboratory Of Wisconsin (GLOW) grid infrastructure at the University of Wisconsin - Madison, the Open Science Grid (OSG) grid infrastructure;
U.S. Department of Energy, and National Energy Research Scientific Computing Center,
the Louisiana Optical Network Initiative (LONI) grid computing resources;
Natural Sciences and Engineering Research Council of Canada,
WestGrid and Compute/Calcul Canada;
Swedish Research Council,
Swedish Polar Research Secretariat,
Swedish National Infrastructure for Computing (SNIC),
and Knut and Alice Wallenberg Foundation, Sweden;
German Ministry for Education and Research (BMBF),
Deutsche Forschungsgemeinschaft (DFG),
Helmholtz Alliance for Astroparticle Physics (HAP),
Research Department of Plasmas with Complex Interactions (Bochum), Germany;
Fund for Scientific Research (FNRS-FWO),
FWO Odysseus programme,
Flanders Institute to encourage scientific and technological research in industry (IWT),
Belgian Federal Science Policy Office (Belspo);
University of Oxford, United Kingdom;
Marsden Fund, New Zealand;
Australian Research Council;
Japan Society for Promotion of Science (JSPS);
the Swiss National Science Foundation (SNSF), Switzerland;
National Research Foundation of Korea (NRF);
Villum Fonden, Danish National Research Foundation (DNRF), Denmark
\end{acknowledgements}

\end{document}